\documentclass[twocolumn,pra,showpacs,aps,superscriptaddress]{revtex4}

\usepackage{amsmath}
\usepackage{amssymb,amsthm}
\usepackage{latexsym}
\usepackage{graphicx}
\usepackage[dvips]{color}
\usepackage{bm}
\usepackage{mathptmx}
\usepackage{times}

\newcommand{\copa}[1]{\ensuremath{\hat{#1}}}
\newcommand{\nbop}[1]{\ensuremath{\copa{n}_{#1}}}

\newcommand{\bd}[1]{\ensuremath{\copa{b}^\dagger_{#1}}}
\newcommand{\bb}[1]{\ensuremath{\copa{b}_{#1}}}

\newcommand{\Exp}[1]{\ensuremath{\text{e}^{#1}}}

\newcommand{\bwave}{\ensuremath{\vert\Psi(\tau)\rangle}}
\newcommand{\biwave}{\ensuremath{\vert\Psi_I\rangle}}
\newcommand{\bfwave}{\ensuremath{\langle\Psi_I\vert}}

\newcommand{\Hop}{\ensuremath{\hat{H}}}
\newcommand{\Nop}{\ensuremath{\hat{N}}}
\newcommand{\abs}[1]{\ensuremath{\left|{#1}\right|}}
\newcommand{\oavg}{\ensuremath{O_j^{\text{av}}}}

\newcommand{\Iop}[1]{\ensuremath{\hat{I}_{#1}}}
\newcommand{\DGop}{\ensuremath{\hat{\rho}_\text{GGE}}}
\newcommand{\DCop}{\ensuremath{\hat{\rho}_\text{GE}}}
\newcommand{\tr}[1]{\ensuremath{\text{Tr}\left[{#1}\right]}}

\begin{document}

\title{Quenches in a quasi-disordered integrable lattice system:\\
Dynamics and statistical description of observables after relaxation}

\author{Christian Gramsch}
\affiliation{Department of Physics, Georgetown University, Washington, DC 20057, USA}
\affiliation{Theoretical Physics III, Center for Electronic Correlations and Magnetism, 
Institute of Physics, University of Augsburg, 86135 Augsburg, Germany}
\author{Marcos Rigol}
\affiliation{Department of Physics, Georgetown University, Washington, DC 20057, USA}
\affiliation{Physics Department, The Pennsylvania State University, 
104 Davey Laboratory, University Park, Pennsylvania 16802, USA}

\begin{abstract}
We study the dynamics and the resulting state after relaxation in a quasi-disordered integrable lattice 
system after a sudden quench. Specifically, we consider hard-core bosons in an isolated one-dimensional 
geometry in the presence of a quasi-periodic potential whose strength is abruptly changed to take the 
system out of equilibrium. In the delocalized regime, we find that the relaxation dynamics of one-body 
observables, such as the density, the momentum distribution function, and the occupation of the natural 
orbitals, follow, to a good approximation, power laws. In that regime, we also show that the observables 
after relaxation can be described by the generalized Gibbs ensemble, while such a description fails for 
the momentum distribution and the natural orbital occupations in the presence of localization. 
At the critical point, the relaxation dynamics is found to be slower than in the delocalized phase.
\end{abstract}

\pacs{05.70.Ln, 72.15.Rn, 02.30.Ik, 05.60.Gg, 71.30.+h}

\maketitle

\section{Introduction}

The nonequilibrium dynamics of isolated integrable quantum systems is constrained by a large number of 
conserved quantities, which generally preclude relaxation to thermal equilibrium 
\cite{rigol_dunjko_07_27,rigol_muramatsu_06_26,cassidy_clark_11_56,cazalilla_06,calabrese_cardy_07a,
cramer_dawson_08,barthel_schollwock_08,eckstein_kollar_08,kollar_eckstein_08,iucci_cazalilla_09,
rossini_silva_09,rossini_susuki_10,fioretto_mussardo_10,iucci_cazalilla_10,mossel_caux_10,
calabrese_essler_11,cazalilla_iucci_12,rigol_fitzpatrick_61,cazalilla_preprint_12,imambekov_preprint_12,
he_rigol_12,calabrese_essler_12a,calabrese_essler_12b}. This may affect current experiments that are 
realized in one-dimensional (1D) and quasi-1D geometries close to integrable points 
\cite{cazalilla_review_11} and future technological devices. As such, this phenomenon cannot be considered 
as purely academic anymore. Advances in controlling and manipulating highly isolated quantum gases in low 
dimensions, and at very low temperatures, have made it possible to study in great detail the relaxation 
dynamics following an abrupt change of some of the system's parameters 
\cite{KinoshitaNature06,trotzky_chen_12}, so that questions related to the lack of thermalization can now 
also be addressed experimentally. For example, in Ref.~\cite{KinoshitaNature06}, it was experimentally 
shown that the relaxation dynamics of one-dimensional atomic Bose gases does not necessarily lead to a 
thermal momentum distribution of the atoms. 

Soon after the experimental finding in Ref.~\cite{KinoshitaNature06}, it was shown in 
Ref.~\cite{rigol_dunjko_07_27} that expectation values of few-body observables in isolated integrable 
systems after relaxation can be predicted by generalized Gibbs ensembles (GGEs). GGEs are constructed by 
maximizing the entropy \cite{jaynes_57a,jaynes_57b}, while satisfying constraints imposed by the constants 
of motion that make the system integrable. Interestingly, the mechanism that leads to thermalization 
in non-integrable systems, namely, eigenstate thermalization 
\cite{DeutschPRA91,SrednickiPRE94,rigol_dunjko_08_34,rigol_srednicki_12_70}, can be generalized to the 
integrable case in the sense that most eigenstates that are close not only in energy but also in their 
distribution of conserved quantities share the same expectation values of few-body observables 
\cite{cassidy_clark_11_56}. This allows one to understand why the GGE description works. Its validity 
after relaxation has been tested in many different integrable quantum models 
\cite{rigol_dunjko_07_27,rigol_muramatsu_06_26,cassidy_clark_11_56,cazalilla_06,calabrese_cardy_07a,
cramer_dawson_08,barthel_schollwock_08,eckstein_kollar_08,kollar_eckstein_08,iucci_cazalilla_09,
rossini_silva_09,rossini_susuki_10,fioretto_mussardo_10,iucci_cazalilla_10,mossel_caux_10,
calabrese_essler_11,cazalilla_iucci_12,rigol_fitzpatrick_61,cazalilla_preprint_12,imambekov_preprint_12,
he_rigol_12,calabrese_essler_12a,calabrese_essler_12b}, and has been argued to be adequate for predicting 
prethermalized expectation values of observables 
\cite{berges_borsanyi_04,moeckel_kehrein_08,moeckel_kehrein_09} in non-integrable quantum systems 
close to an integrable point \cite{KollarPRB11}.

In relation to current ultracold-gas experiments (and to low-dimensional mesoscopic devices), one 
question that needs to be addressed is the fate of the GGE description when translational invariance 
is absent in the system. Numerical calculations for hard-core bosons in a box 
\cite{rigol_dunjko_07_27,cassidy_clark_11_56,rigol_muramatsu_06_26} and in the presence of a harmonic 
confining potential (relevant to optical lattice setups) \cite{rigol_muramatsu_06_26,cazalilla_iucci_12}, 
have shown that the GGE indeed describes observables after relaxation. However, a recent study of quenches 
in the quantum Ising chain has put forward the notion that ``as soon as the translational invariance is 
broken, the GGE fails to apply'' \cite{caneva_canovi_11}. This was supported by calculations of equal-time 
correlations after a quench in the presence of disorder. Since the general statement made in 
Ref.~\cite{caneva_canovi_11} is in contradiction with previous results 
\cite{rigol_dunjko_07_27,cassidy_clark_11_56,rigol_muramatsu_06_26,cazalilla_iucci_12}, especially 
with those in the presence of a confining potential \cite{rigol_muramatsu_06_26,cazalilla_iucci_12}, 
here we reconsider the question of whether the GGE description is valid in the absence of translational invariance.

One important difference between the systems studied in Ref.~\cite{caneva_canovi_11} and those studied 
in Refs.~\cite{rigol_dunjko_07_27,cassidy_clark_11_56,rigol_muramatsu_06_26} is the inclusion of disorder 
in the former. Even in the presence of interactions, disorder can lead to localization 
\cite{basko_alainer_06,oganesyan_huse_07,pal_huse_10,khatami_rigol_12}, and, in nonintegrable systems,
localization can lead to lack of thermalization after relaxation following a quantum quench 
\cite{khatami_rigol_12,gogolin_muller_11}. The latter can be understood to follow from the failure 
of eigenstate thermalization in the localized regime \cite{khatami_rigol_12}. It is then natural 
to expect that, in integrable systems, localization, and not necessarily the breaking of translational 
symmetry, may lead to a failure of the GGE description. This would follow from a failure of the generalized eigenstate 
thermalization \cite{cassidy_clark_11_56}. 

In order to separate the effects of breaking translational symmetry and localization in an integrable 
system, we study hard-core bosons in an incommensurate superlattice. This model exhibits a transition
between an extended and a localized phase at a finite strength of the superlattice potential 
\cite{aubry_andre_80}, and is to be contrasted with the case of uniform random disorder where localization 
occurs for any nonzero disorder strength \cite{cazalilla_review_11}. We show that in the extended phase, 
the GGE provides a correct description of one-body observables after relaxation, despite the lack of 
translational invariance. On the other hand, in the localized phase, the GGE is found to fail. At the 
critical point, a slower relaxation dynamics is seen to preclude the observation of stationary values 
of the observables for the largest system sizes. However, as long as the stationary value is reached, 
the GGE provides a good description of observables after relaxation at the critical point.

The exposition is organized as follows. In the next section (Sec.~\ref{sec:intModel}), we introduce the 
model and observables to be studied in the remainder of the paper. We also briefly discuss the computational 
approach utilized in our study, as well as the ensembles that are used to compare with the results after 
relaxation. In Sec.~\ref{sec:relax}, we study the relaxation dynamics following a sudden quench in the 
different regimes of the model. Section \ref{sec:steadystate} is devoted to a comparison of observables 
after relaxation with the predictions of statistical ensembles, as well as a finite-size-scaling analysis 
that allows us to gain insight into the behavior in the thermodynamic limit. We also make contact 
with the results in Ref.~\cite{caneva_canovi_11} by studying the behavior of off-diagonal 
one-particle correlations. The conclusions are presented in Sec.~\ref{sec:summary}.

\section{Model, observables, and ensembles}\label{sec:intModel}

Our study is performed within the Aubry-Andr\'e model \cite{aubry_andre_80} for hard-core bosons in a 
one-dimensional lattice with open boundary conditions. The Hamiltonian reads 
\begin{equation}
\Hop=-t\sum_{j=1}^{L-1} \left(\bd{j} \bb{j+1} + \text{H.c.}\right) 
+ \lambda\sum_{j=1}^L \cos\left(2\pi\sigma j+\varphi\right)\nbop{j},
\label{eq:Hamiltonian}
\end{equation}
where the operator $\hat{b}_j^\dagger\,\,(\hat{b}^{}_j)$ creates (annihilates) a hard-core boson at site $j$, 
and $\hat{n}_j=\hat{b}_j^\dagger\hat{b}^{}_j$ is the on-site occupation number operator. $\hat{b}^{}_j$ and 
$\hat{b}_j^\dagger$ obey the usual bosonic commutation relations, i.e., 
$[\hat{b}^{}_i,\hat{b}_j^\dagger]=\delta_{ij}$, but satisfy a constraint 
$\hat{b}_j^2=\hat{b}_j^{\dagger2}=0$, which forbids multiple occupancy of the lattice sites. The hopping 
parameter is denoted by $t$ (we set $t=1$, $\hbar=1$ throughout this work), $L$ is the number of sites,
and we consider only systems in which the number of particles ($N$) is $N=L/2$ (half filling). 
By selecting $\sigma$ to be an irrational number, we generate a quasiperiodic potential whose strength is 
controlled by the parameter $\lambda$. In our study, we choose $\sigma$ to be the inverse golden ratio,
$\sigma=(\sqrt{5}-1)/2$, a choice motivated by the fact that the golden mean is considered to be the most 
irrational number \cite{sokoloff_review_85}. $\varphi$ allows 
the phase of the potential to be shifted, and will be used later to average over different realizations in our
finite systems. For most of our work, we set $\varphi=0$.

Despite the quadratic form of Eq.~\eqref{eq:Hamiltonian}, it cannot be directly diagonalized because 
of the on-site constraints forbidding multiple occupancy of the lattice sites. This can, however, be 
circumvented by mapping the 1D hard-core boson Hamiltonian onto a spin-1/2 chain via the Holstein-Primakoff 
transformation \cite{hp:12.40}, and then mapping the spin-1/2 chain onto noninteracting spinless fermions 
\cite{lieb_shultz_61} via the Jordan-Wigner transformation \cite{jw:28}. The resulting Hamiltonian maintains 
the form in Eq.~\eqref{eq:Hamiltonian} but with the hard-core operators replaced by fermionic ones. It then 
follows that the spectrum, as well as thermodynamic and density-related properties, are the same for 
hard-core bosons and non-interacting spinless fermions.

The Aubry-Andr\'e model \cite{aubry_andre_80} is known to undergo a localization transition at a critical 
$\lambda_c=2t$.  For $\lambda <\lambda_c$, all single-particle states are extended, i.e., Bloch-like, states. 
Above the critical point, single-particle states are exponentially localized with localization length 
$\xi = \ln(\lambda)^{-1}$ \cite{aubry_andre_80}. Because of the mapping above, the same holds true for 
hard-core bosons. This implies that the ground state of the latter undergoes a superfluid-insulating 
transition as $\lambda_c=2t$ is crossed. In the localized phase, the ground state is a Bose glass 
\cite{cazalilla_review_11}.

In connection to optical lattice experiments, such as the ones carried out in 
Refs.~\cite{KinoshitaNature06,trotzky_chen_12}, we are interested in studying two different one-body 
observables: the on-site density $n_i=\langle \hat{n}_i \rangle$, and the momentum distribution function 
$m_k$. $m_k$ is the diagonal part of the Fourier transform of the one-particle density matrix 
$\rho_{ij}=\langle \hat{b}_i^\dagger\hat{b}^{}_j \rangle$,
\begin{equation}
m_k=\frac{1}{L}\sum_{i,j=1}^L{\Exp{\text{i} k(i-j)}}\, \rho_{ij}.
\end{equation}
Additional information on the coherence properties of the system can be gained through the study of the 
natural orbitals $\phi_i^\alpha$ and their occupations $\eta_\alpha$, defined through the eigenvalue 
equation
\begin{equation}
	\sum_{j=1}^L\rho_{ij}\,\phi^\alpha_j=\eta_\alpha \,\phi^\alpha_i.
\end{equation}
In homogeneous periodic systems, the natural orbitals are plane waves and their occupations 
coincide with the momentum distribution function, so $m_k$ and $\eta_\alpha$ give the same information 
about the system. However, once translational invariance is broken these two quantities become 
different. Out of equilibrium, they can even give apparently inconsistent results. For example, 
during the expansion of a hard-core boson gas its momentum distribution function becomes identical 
to that of noninteracting fermions, which may be taken as an indication that the system lacks coherence 
\cite{rigol_muramatsu_05_12}. However, the occupation of the natural orbitals is very different from 
the one of fermions; many orbitals remain highly populated, which reveals the bosonic character of the 
out-of-equilibrium gas \cite{rigol_muramatsu_05_12}. In addition, in higher-dimensional interacting systems, 
if the occupation of the highest occupied natural orbital scales with the total number of particles, 
then one can say that the system exhibits Bose-Einstein condensation \cite{po:11.56,leggett_review_01}.

In equilibrium, the properties of hard-core bosons, modeled by Eq.~\eqref{eq:Hamiltonian}, have been
studied in detail in the ground state \cite{rey_satija_06b,he_satija_12_67} and at finite temperature
\cite{nessi_iucci_11}. Here, our goal is to examine the dynamics after the system is taken out of 
equilibrium by a sudden change of $\lambda$ ($\lambda_I\rightarrow\lambda_F$). The initial state 
$\biwave$ is taken to be the ground state of $\Hop_I$ [Eq.~\eqref{eq:Hamiltonian} with $\lambda=\lambda_I$] 
and the evolution is studied under $\Hop_F$ [Eq.~\eqref{eq:Hamiltonian} with $\lambda=\lambda_F$]:
\begin{equation}
	\bwave =\Exp{-\text{i}\Hop_{F}\tau}\biwave.
\end{equation}

To study the time evolution of the observables introduced above, we follow a computational method based 
on the Bose-Fermi mapping and the use of properties of Slater determinants. This method 
has been explained in detail in Ref.~\cite{rigol_muramatsu_05_16}, so we do not reproduce it here. It
allows one to calculate each matrix element $\rho_{ij}$ (at any given time $\tau$) in terms of the 
determinant of an $(N+1)\times(N+1)$ matrix, which results from the product of two matrices with sizes 
$(N+1)\times L$ and $L\times(N+1)$. The computation time of the entire one-particle density matrix 
essentially scales as $L^2(N + 1)^3$ (the matrix multiplication need not be done for every entry), 
which allows us to efficiently study the dynamics of systems of up to 1000 lattice sites.

We then contrast the time-averaged results for the observables after relaxation with the predictions of 
statistical mechanics. While the most relevant traditional ensemble to compare with would be the 
microcanonical one (because the time-evolving system is isolated), we instead use the grand-canonical 
ensemble (GE). This is because calculations in the former scale exponentially with system size, while, 
in the latter, they scale as power laws. Within the GE, we can study very large 
lattices, in which we expect a good agreement between the predictions from different 
statistical ensembles \cite{rigol_05_19}. The density matrix in the GE reads
\begin{equation}
	\DCop=\frac{1}{Z_\text{GE}}\exp\left(-\frac{\Hop-\mu\Nop}{k_B T}\right),
	\label{eq:Definition GE}
\end{equation}
where $k_B$ is the Boltzmann constant, $\hat{N}$ is the total number operator, 
and $Z_\text{GE}$ is the partition function,
\begin{equation}
	Z_\text{GE}=\tr{\exp\left(-\frac{\Hop-\mu\Nop}{k_B T}\right)}.
	\label{eq:normalization Z_GE}
\end{equation}
In order to compare the grand-canonical predictions for the observables to those obtained following 
the quantum dynamics, $T_{}$ and $\mu$ need to be chosen so that 
$\textrm{Tr}[\hat\rho_{\textrm{GE}}\hat H_F]=E$ and $\textrm{Tr}[\hat\rho_{\textrm{GE}} \hat N]=N$,
where $E=\bfwave\hat{H}_F\biwave$ is the energy of the time-evolving system after the quench, which 
is conserved. 

In integrable hard-core-boson systems, in the absence of disorder or quasi-disorder, the grand-canonical 
\cite{rigol_dunjko_07_27,cassidy_clark_11_56,rigol_muramatsu_06_26} and microcanonical 
\cite{cassidy_clark_11_56} descriptions have been shown to fail to predict the outcome of the relaxation 
dynamics for few-body observables. Instead, the GGE has been proposed to be the adequate ensemble to deal 
with this problem \cite{rigol_dunjko_07_27}. The GGE density matrix can be written as
\begin{equation}
	\DGop=\frac{1}{Z_\text{GGE}}\exp\left(-\sum_{m}\lambda_m\Iop{m}\right),
	\label{eq:Definition GGE}
\end{equation}
where $\Iop{m}$ are the conserved quantities, $\lambda_m$ are their corresponding Lagrange multipliers, and 
$Z_\text{GGE}$ is the partition function,
\begin{equation}
	Z_\text{GGE}=\tr{\exp\left(-\sum_{m}\lambda_m\Iop{m}\right)}.
	\label{eq:normalization Z_GGE}
\end{equation}
The Lagrange multipliers need to be selected so that the expectation values of the conserved quantities
in the GGE are the same as in the initial state, i.e.,
$\textrm{Tr}[\hat\rho_{\textrm{GGE}} \hat I_m]=\bfwave\hat{I}_m\biwave$.
For hard-core bosons, where the conserved quantities are taken to be the projection operators to the 
single-particle eigenstates of the fermionic Hamiltonian to which Eq.~\eqref{eq:Hamiltonian} 
can be mapped, the Lagrange multipliers can be written as \cite{rigol_dunjko_07_27}
 \begin{equation}
\lambda_m=\ln \left(\frac{1-\bfwave\hat{I}_m\biwave}{\bfwave\hat{I}_m\biwave}\right).
\label{eq:lagrangem}
 \end{equation}

In order to calculate the expectation value of the one-particle density matrix in the grand-canonical
ensemble, $\rho_{ij}^{\text{GE}}=\tr{\hat{b}_i^\dagger\hat{b}^{}_j \hat\rho_{\textrm{GE}}}$, and in the GGE, 
$\rho_{ij}^{\text{GGE}}=\tr{\hat{b}_i^\dagger\hat{b}^{}_j \hat\rho_{\textrm{GGE}}}$ (note that the GGE is 
also grand-canonical), we use the approach introduced in Ref.~\cite{rigol_05_19}. The grand-canonical 
calculations, similarly to the ones carried out for studying the dynamics, use the Bose-Fermi mapping and 
properties of Slater determinants. The computation time of the entire one-particle density matrix in 
this case scales as $L^5$ \cite{rigol_05_19}.

\section{Time Evolution}\label{sec:relax}

To probe the relaxation dynamics after the quench, we calculate the normalized difference 
$\delta O(\tau)$ (where $O$ stands for $n$, $m$, $\eta$) between the expectation value of observables 
at different times and their long-time average \oavg. $\delta O(\tau)$ is 
defined as
\begin{equation} \label{eq:deltaO}
	\delta O(\tau)=\frac{\sum_{j}\abs{O_j(\tau)-\oavg}}{\sum_{j}\oavg}.
\end{equation}
[Note that $j$ is a dummy variable that stands for $i$ (in $n_i$), $k$ (in $m_k$), and $\alpha$ (in $\eta_\alpha$)].
If observables relax to stationary values, $\delta O(\tau)$ will fluctuate about a time-independent 
value. This value, as well as the amplitude of the time fluctuations about it, is expected to be finite 
for finite systems but should vanish in the thermodynamic limit. We note that $\oavg$ is taken to be an 
average over a variable size time interval that contains the longest times that we have simulated. In 
the event that the observable has not relaxed by then, $\delta O(\tau)$ will make it evident as it will 
not become stationary.

\begin{figure*}[!t]
  \centering
  \includegraphics[width=0.98\textwidth]{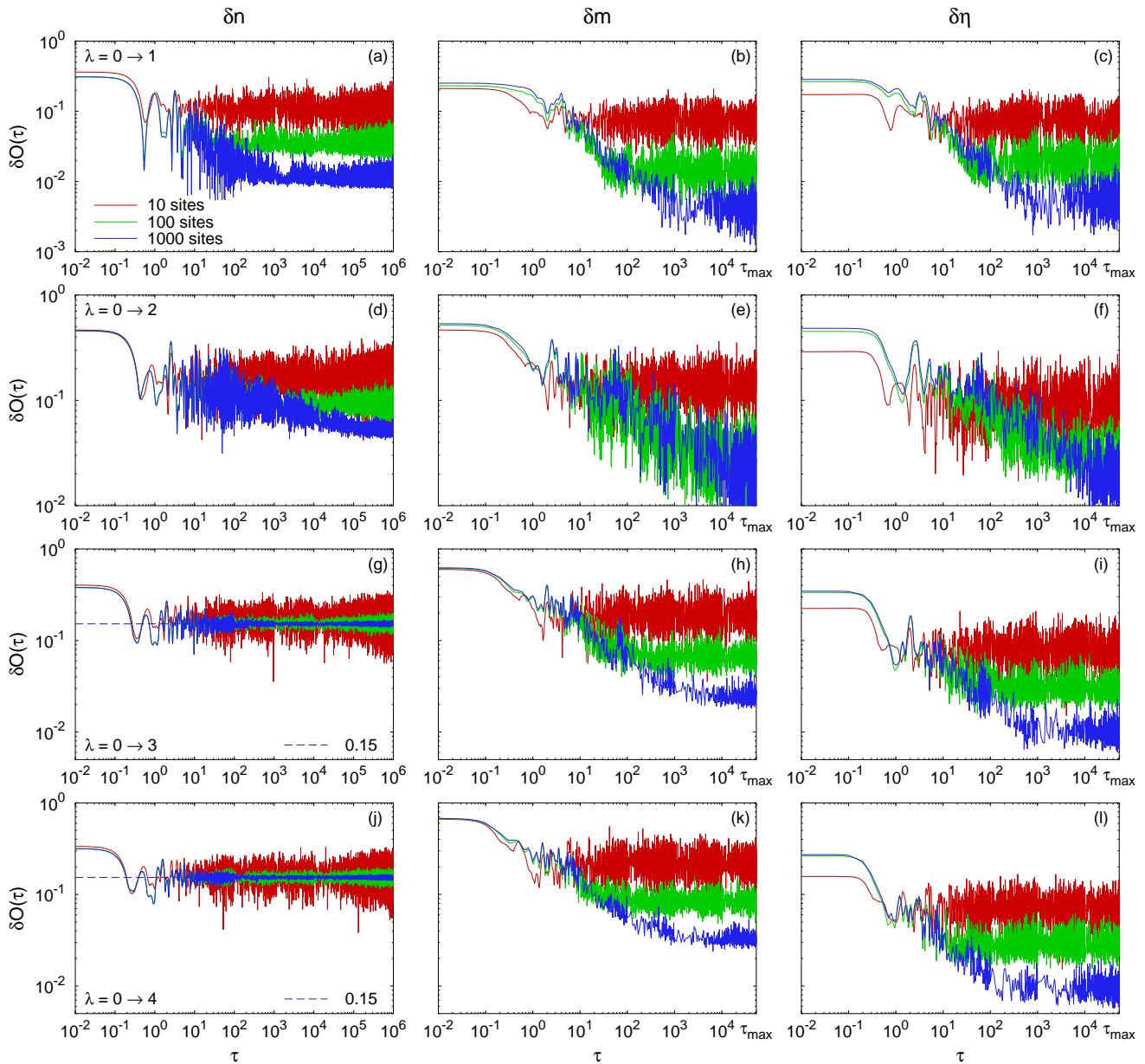}
  \vspace{-0.55cm}
  \caption{(Color online) Relaxation dynamics of $n_i$, $m_k$, and $\eta_\alpha$ 
as they approach the long-time average in a quench $\lambda_I=0\rightarrow\lambda_F=1$
(a)--(c), $\lambda_I=0\rightarrow\lambda_F=2$ (d)--(f), $\lambda_I=0\rightarrow\lambda_F=3$ (g)--(i),
and $\lambda_I=0\rightarrow\lambda_F=4$ (j)--(l), for systems with 10, 100, and 1000 lattice sites 
(from top to bottom in each panel). The time averages are computed as follows: Since $n_i$ is 
computationally less expensive than $m_k$ and $\eta_\alpha$, for that observable we simulated longer times 
and averaged over $9000$ steps with $\tau\in[10^5,10^6]$ for all lattice sizes. For $m_k$ and $\eta_\alpha$, 
we averaged over $900$ steps with $\tau\in[10^4,10^5]$ for $L=10$ and $L=100$, and over 437 steps for 
$L=1000$ with $\tau\in[10^4,5.37\times 10^4]$ (in the plots, $\tau_\text{max}=5.37\times 10^4$). }
\label{fig:relaxnk}	
\end{figure*}

In Fig.~\ref{fig:relaxnk}, we show results for $\delta O(\tau)$ in a set of quenches in which the initial 
state is the ground state of Eq.~\eqref{eq:Hamiltonian} with $\lambda_I=0$ (i.e., a superfluid state) 
and $\lambda_F$ is below ($\lambda_F=1$), at ($\lambda_F=2$), and above ($\lambda_F=3,4$) the localization 
transition. Results are presented for three different system sizes ($L=10,\,100,$ and 1000, from top to 
bottom in each panel). In Figs.~\ref{fig:relaxnk}(a)--\ref{fig:relaxnk}(c), one can see that all three 
observables in the quench terminating in the extended phase exhibit a clear relaxation dynamics in which 
$\delta O(\tau)$ decreases as time passes, and then fluctuates about a finite time-independent value. 
Both the finite time-independent value and the amplitude of the fluctuations are seen to decrease with 
increasing system size.

The quench towards the critical point ($\lambda_F=2$) [Figs.~\ref{fig:relaxnk}(d)--\ref{fig:relaxnk}(f)] 
exhibits a different dynamics. As the system size increases beyond 100 sites, the three observables 
considered here do not reach a clear stationary value during the times studied (up to $\tau=10^6$ for 
$n_i$ and $\tau=5.37\times10^4$ for $m_k$ and $\eta_\alpha$). This can be understood as the critical point is known 
to be very special. The single-particle spectrum becomes a Cantor set (the bands acquire zero measure), 
and the gaps form a devil's staircase \cite{harper}. Such a peculiar spectrum seems to render dephasing 
ineffective in these systems. Our finding implies that, at the critical point, stationary values of the 
observables may be more difficult to observe experimentally.

Finally, the quench towards the localized regime  [Figs.~\ref{fig:relaxnk}(g)--\ref{fig:relaxnk}(i) 
for $\lambda_F=3$ and Figs.~\ref{fig:relaxnk}(j)--\ref{fig:relaxnk}(l) for $\lambda_F=4$] does 
lead to stationary values for $m_k$ and $\eta_\alpha$. Note that $m_k$ and $\eta_\alpha$ exhibit 
dynamics that are qualitatively similar to the one observed in the quench 
$\lambda_I=0\rightarrow\lambda_F=1$, namely, the stationary values of $\delta m$ and $\delta \eta$ 
(and the fluctuations about them) decrease with increasing system size. $n_i$, on the other hand, 
exhibits a different behavior. Because of localization in real space, $\delta n$ becomes lattice-size 
independent, i.e., it remains finite in the thermodynamic limit. In that case, the only effect 
that increasing $L$ has is to reduce the amplitude of the time fluctuations of $\delta n$ about the 
stationary value.

\begin{figure*}[!ht]
  \centering
  \includegraphics[width=0.98\textwidth]{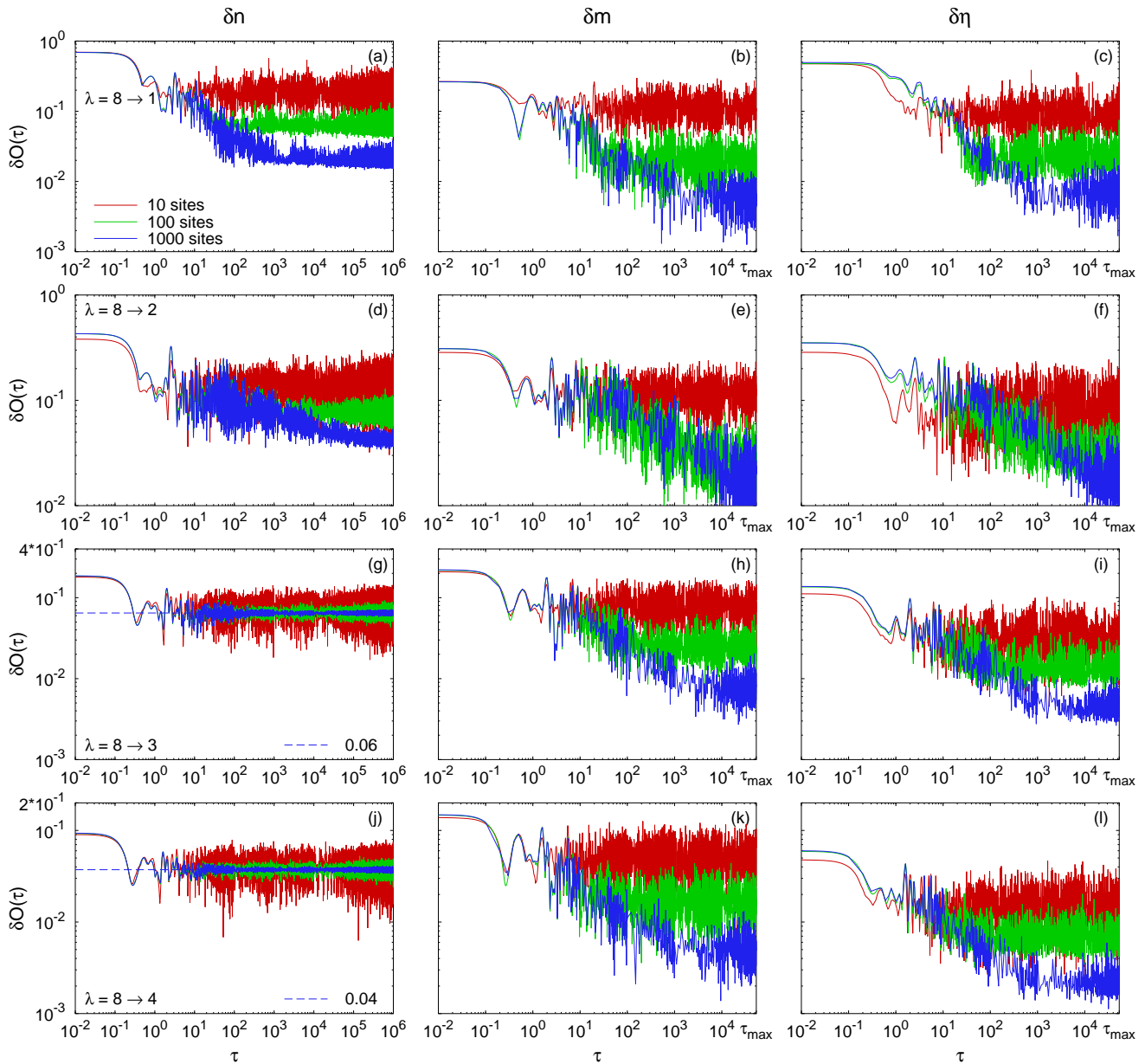}
  \vspace{-0.55cm}
  \caption{(Color online) As Fig. \ref{fig:relaxnk} but for quenches from $\lambda_I=8$, i.e., from 
deep inside the Bose-glass phase.}
	\label{fig:relaxnk2}
\end{figure*}

We have also studied quenches starting from different initial states that are eigenstates of 
Eq.~\eqref{eq:Hamiltonian}, and even from the ground state of commensurate superlattices such as the ones studied 
in Refs.~\cite{rigol_muramatsu_06_26,rigol_fitzpatrick_61,he_rigol_12}, finding a qualitatively similar 
dynamics to the one depicted in Fig.~\ref{fig:relaxnk}. As an example of a different initial state, in 
Fig.~\ref{fig:relaxnk2}, we report results in which the quenches start from the ground state of Hamiltonian 
\eqref{eq:Hamiltonian} deep inside the Bose-glass phase ($\lambda_I=8$). Figure \ref{fig:relaxnk2} shows 
that the dynamics is indeed very similar to that reported in Fig.~\ref{fig:relaxnk}. The only apparent 
difference is that for quenches within the Bose-glass phase ($\lambda_I=8\rightarrow\lambda_F=3$ and 
$\lambda_I=8\rightarrow\lambda_F=4$), the stationary value of $\delta n$ is smaller than in the quenches 
from the superfluid phase to the Bose-glass phase ($\lambda_I=0\rightarrow\lambda_F=3$ and 
$\lambda_I=0\rightarrow\lambda_F=4$). For the former, we find 
$\delta {n}^{8\rightarrow3}(\infty)\approx0.06$ and $\delta{n}^{8\rightarrow4}(\infty)\approx0.04$ while 
for the latter $\delta {n}^{0\rightarrow3}(\infty)\approx\delta {n}^{0\rightarrow4}(\infty)\approx0.15$.
This is understandable as $\delta n(0)$ is already smaller in quenches starting in the Bose-glass phase 
than in those starting in the superfluid phase.

\subsection*{Approach to the stationary values}

In a recent numerical study of the relaxation dynamics of a disordered nonintegrable fermionic 
system with short-range interactions and random long-range hopping, it was found that, in the 
extended phase, observables exhibit a power-law approach to their thermal expectation values 
\cite{khatami_rigol_12}. Power-law-like relaxation dynamics was also seen in recent optical 
lattice experiments with a clean system in a one-dimensional geometry \cite{trotzky_chen_12}. 
These results are to be contrasted with the exponential approach expected 
in generic nonintegrable systems. Since both studies 
\cite{khatami_rigol_12,trotzky_chen_12} were limited to small lattice sizes, 
and no extensive scaling analysis could be performed, it is not clear how these findings are 
affected by finite-size effects.

\begin{figure}[!t]
  \includegraphics[width=0.48\textwidth]{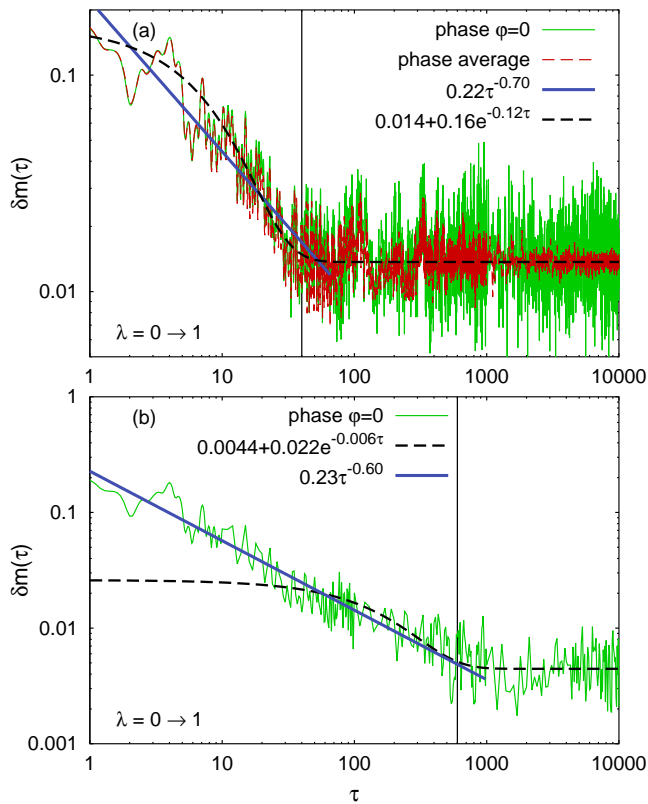}
  \vspace{-0.3cm}
  \caption{(Color online) (a) $\delta m_k$ vs $\tau$ for $\varphi=0$, as well as after averaging over 
$1000$ random values of $\varphi$ (uniformly distributed in $[0,2\pi]$), in systems with 100 lattice 
sites. The fits to power-law and exponential behavior were done over the interval $\tau\in[1,40]$ 
(a vertical line marks $\tau=40$), which contains $1200$ data points. (b) $\delta m_k$ vs $\tau$ 
for $\varphi=0$ in a system with 1000 lattice sites. The fits to power-law and exponential behavior 
were done over the interval $\tau\in[1,600]$ (a vertical line marks $\tau=600$), which contains 
230 data points.}\label{fig:powerrelax}
\end{figure}

The dynamics depicted in Figs.~\ref{fig:relaxnk} and \ref{fig:relaxnk2} for three system sizes, 
which are a decade away from each other, provides a clearer picture of the role of finite-size 
effects. We indeed find indications of power-law relaxation, as it is apparent in the plots 
that the time interval over which a power-law-like behavior is seen increases with system size. 
We explicitly show this in Fig.~\ref{fig:powerrelax}, where we compare the relaxation process for 
systems with 100 and 1000 lattice sites. In the former [Fig.~\ref{fig:powerrelax}(a)], both power-law 
and exponential decay provide a reasonable fit to the data. In the latter [Fig.~\ref{fig:powerrelax}(b)], 
where power-law behavior is apparent for about three decades, a fit to an exponential decay is clearly 
inconsistent with the data. Hence, our results provide another example of a system in which, whenever 
relaxation takes place, the relaxation dynamics follows a power law. To what extent power-law-like relaxation 
is generic to the dynamics of isolated quantum systems, especially nonintegrable ones, is a topic that 
deserves further attention.

Since we are dealing with finite lattice sizes with open boundary 
conditions, we have also studied the effect that averaging over different phases 
$\varphi$ [see Eq.~\eqref{eq:Hamiltonian}] has on our results. A typical outcome of 
such an average is depicted in Fig.~\ref{fig:powerrelax}(a), for $1000$ different values of 
$\varphi$ distributed uniformly in $[0,2\pi]$. The average over different phases 
can be seen to reduce time fluctuations after relaxation, but leaves the results 
for the approach to the stationary value almost unaffected. 

\begin{figure}[!t]
	\centering
	\includegraphics[width=0.48\textwidth]{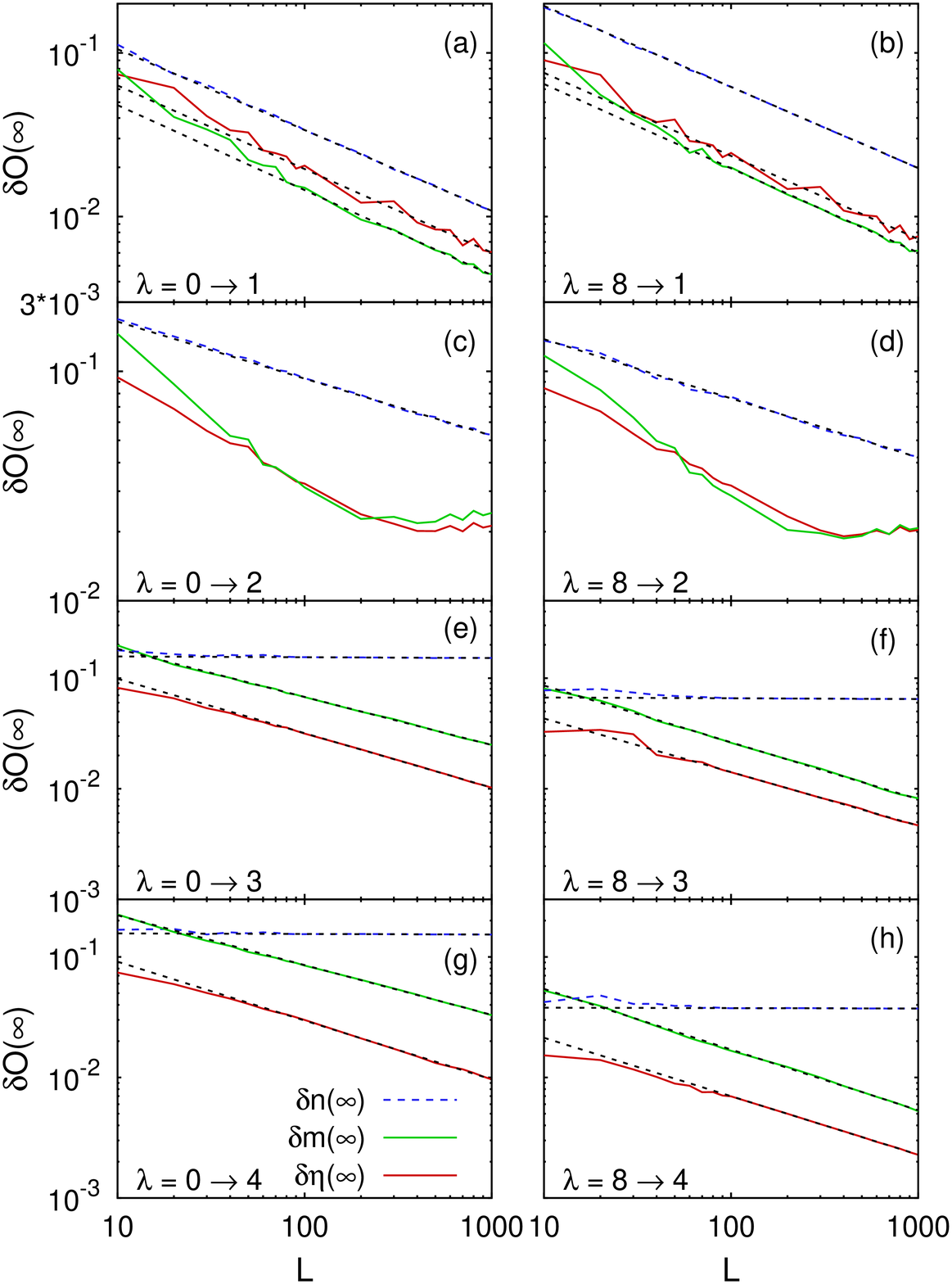}
  \vspace{-0.3cm}
	\caption{(Color online) Finite-size scaling of $\delta n(\infty)$, 
$\delta m(\infty)$, and $\delta \eta(\infty)$ for the quenches studied in 
Figs.~\ref{fig:relaxnk} and \ref{fig:relaxnk2}. The dashed lines are power-law fits leading to 
$\delta n(\infty)\propto L^{-0.49}$, $\delta m(\infty)\propto L^{-0.52}$, and 
$\delta \eta(\infty)\propto L^{-0.51}$ in (a), 
$\delta n(\infty)\propto L^{-0.50}$, $\delta m(\infty)\propto L^{-0.52}$, and 
$\delta \eta(\infty)\propto L^{-0.51}$ in (b),
$\delta n(\infty)\propto L^{-0.25}$ in (c),
$\delta n(\infty)\propto L^{-0.26}$ in (d),
$\delta n(\infty)\propto L^{-0.01}$, $\delta m(\infty)\propto L^{-0.43}$, and 
$\delta \eta(\infty)\propto L^{-0.49}$ in (e),
$\delta n(\infty)\propto L^{-0.01}$, $\delta m(\infty)\propto L^{-0.51}$, and 
$\delta \eta(\infty)\propto L^{-0.48}$ in (f),
$\delta n(\infty)\propto L^{0}$, $\delta m(\infty)\propto L^{-0.41}$, and 
$\delta \eta(\infty)\propto L^{-0.49}$ in (g), and 
$\delta n(\infty)\propto L^{0}$, $\delta m(\infty)\propto L^{-0.50}$, and 
$\delta \eta(\infty)\propto L^{-0.48}$ in (h).
The power-law fits were done using the data for systems with between 100 and 1000 lattice sites 
(11 data points).}
	\label{fig:Scaling5}
\end{figure}

Another important question to be answered, which is of special interest to current experiments with 
ultracold gases, is how long it takes for observables to reach the stationary values. Given the strong 
indications found above that the relaxation dynamics follows a power law, the times at which stationary values 
are attained will be determined by how $\delta n(\infty)$, $\delta m(\infty)$, and $\delta \eta(\infty)$ 
(here, ``$\infty$'' should be understood as a long time after relaxation) scale with system size. In 
Fig.~\ref{fig:Scaling5}, we show the scaling of those quantities in the 
quenches analyzed in Figs.~\ref{fig:relaxnk} and \ref{fig:relaxnk2}. Figures \ref{fig:relaxnk}(a), 
\ref{fig:relaxnk}(b), and \ref{fig:relaxnk}(e)--\ref{fig:relaxnk}(f) show that, away from the critical 
point, the scaling of $\delta m(\infty)$ and $\delta \eta(\infty)$ is close to $1/\sqrt{L}$, and a 
similar scaling is seen for $\delta n(\infty)$ in quenches to the extended phase 
[Figs.~\ref{fig:relaxnk}(a) and \ref{fig:relaxnk}(b)]. Such a scaling has been proven to provide a 
bound for the normalized time variance of observables that are quadratic in Fermi operators in noninteracting-fermion
models \cite{campos12}, but we find it to be also applicable to more general observables in integrable 
systems. As discussed before, in quenches to the localized regime, $\delta n(\infty)$ becomes independent of 
system size. Also, the slow relaxation dynamics of $m$ and $\eta$ at the critical point 
precludes the observation of a clear scaling for $\delta m(\infty)$ and $\delta \eta(\infty)$ 
[Figs.~\ref{fig:relaxnk}(c) and \ref{fig:relaxnk}(d)], while the scaling of $\delta n(\infty)$ is close 
to $1/L^{1/4}$. The scalings of $\delta n(\infty)$ at the critical point and in the localized
regime violate the bound proven in Ref.~\cite{campos12}.

A power-law approach of $\delta n(\tau)$, $\delta m(\tau)$, and $\delta \eta(\tau)$
to the stationary values, together with a power-law scaling of $\delta n(\infty)$, $\delta m(\infty)$, 
and $\delta \eta(\infty)$ with system size, implies that the time at which stationary values 
are attained increases as a power law with system size. This means that measuring densities and 
momentum distribution functions in experiments is advantageous with respect to directly measuring 
two-point correlation functions. After relaxation, the values of the latter have been shown to be 
exponentially small compared with the distance between the points 
\cite{rigol_muramatsu_06_26,calabrese_essler_12b} and, as such, the time it takes for those 
correlations to relax to the stationary values increases exponentially with the distance between 
the points \cite{calabrese_essler_12b}.

\section{Description after relaxation}\label{sec:steadystate}

After discussing the relaxation dynamics, we focus on the description of the observables after relaxation. 
In generic (non-integrable) quantum systems, one expects the dynamics to lead to thermalization, namely, 
to expectation values of observables that are equal to those of a system in thermal equilibrium. Because 
of thermodynamic universality, this is expected to be true whenever the isolated system and its thermal 
equilibrium counterpart share the same mean energy and number of particles 
\cite{DeutschPRA91,SrednickiPRE94,rigol_dunjko_08_34,rigol_srednicki_12_70}, independently of
the initial state in the former.

\onecolumngrid

\begin{figure*}[!b]
	\centering
	\includegraphics[width=0.98\textwidth]{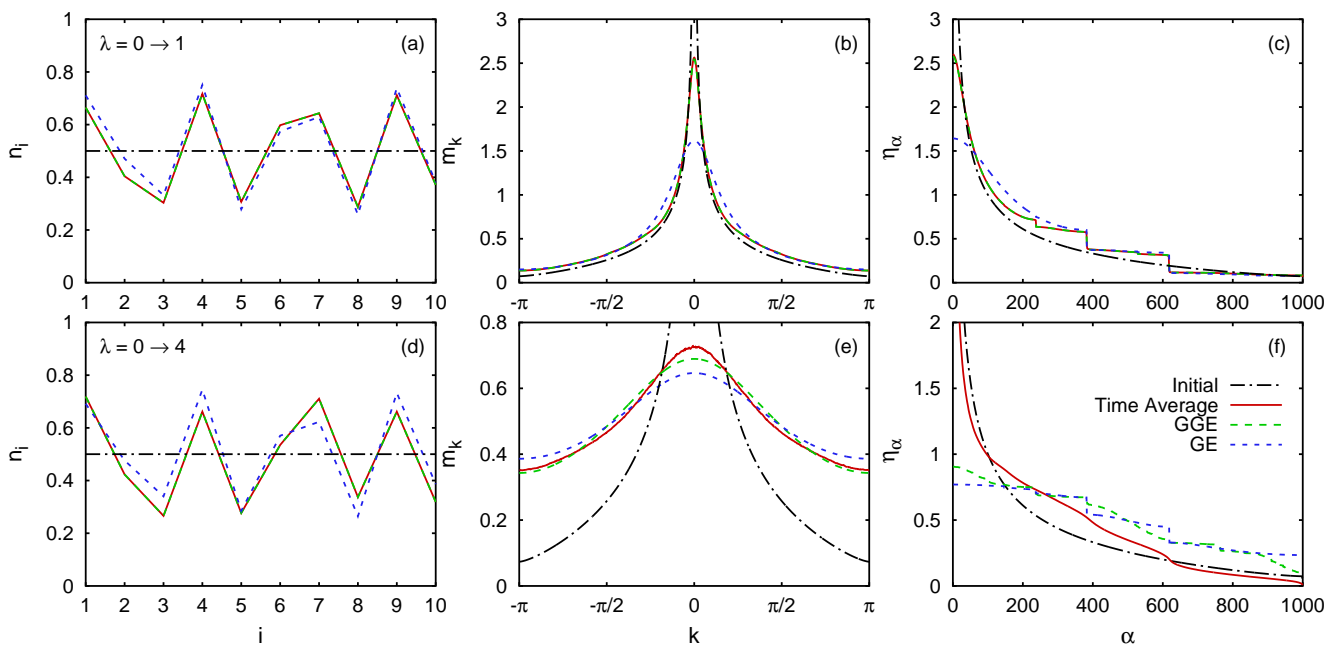}
	\caption{(Color online) Density in the first ten sites (a),(d), momentum distribution function 
(b),(e), and natural orbital occupations (c),(f) for quenches in which the initial state is the 
superfluid ground state of a system with $\lambda_I=0$ while $\lambda_F=1$ (a)--(c), 
$\lambda_F=4$ (d)--(f), and for $L=1000$. We present results for the observables in the initial state, the 
long-time average [calculated between $\tau=10^5$ and $\tau=10^6$ for $n_i$ (9000 steps), and between $\tau=10^4$ 
and $\tau=5.37\times10^4$ (437 steps) for $m_k$ and $\eta_\alpha$; see the caption of Fig.~\ref{fig:relaxnk}], 
as well as within the GE and the GGE. Note that, except $\delta n(\infty)$ for $\lambda_F=4$, 
$\delta n(\infty)$, $\delta m(\infty)$, and $\delta\eta(\infty)$ are very small for $L=1000$ 
(see Fig.~\ref{fig:Scaling5}). In addition, we have checked that all time averages are well converged.}
        \label{fig:lambda=0->1,4}
\end{figure*}

\twocolumngrid

In Fig.~\ref{fig:lambda=0->1,4}, we show results for $n_i$, $m_k$, and $\eta_\alpha$ for quenches from 
initial states with $\lambda_I=0$, and $\lambda_F=1$ and 4. For all quantities, we report their values in 
the initial state, the long-time averages, and within the GE and the GGE. The plots for 
the density in the initial state [Figs.~\ref{fig:lambda=0->1,4}(a) and \ref{fig:lambda=0->1,4}(d)] make 
evident that, despite the presence of open boundary conditions, at $\tau=0$ the density is constant 
throughout the system. This is because of the particle-hole symmetry of the model. After the quench, 
this is not true anymore and the density becomes time dependent and inhomogeneous. The 
time-averaged result for the density after relaxation and the predictions of the GGE are 
indistinguishable from each other for $\lambda_F=1$ in Fig.~\ref{fig:lambda=0->1,4}(a) and 
$\lambda_F=4$ in Fig.~\ref{fig:lambda=0->1,4}(d). The predictions of the GE are different from 
the outcome of the relaxation dynamics in both quenches.

\onecolumngrid

\begin{figure*}[!t]
	\centering
	\includegraphics[width=0.98\textwidth]{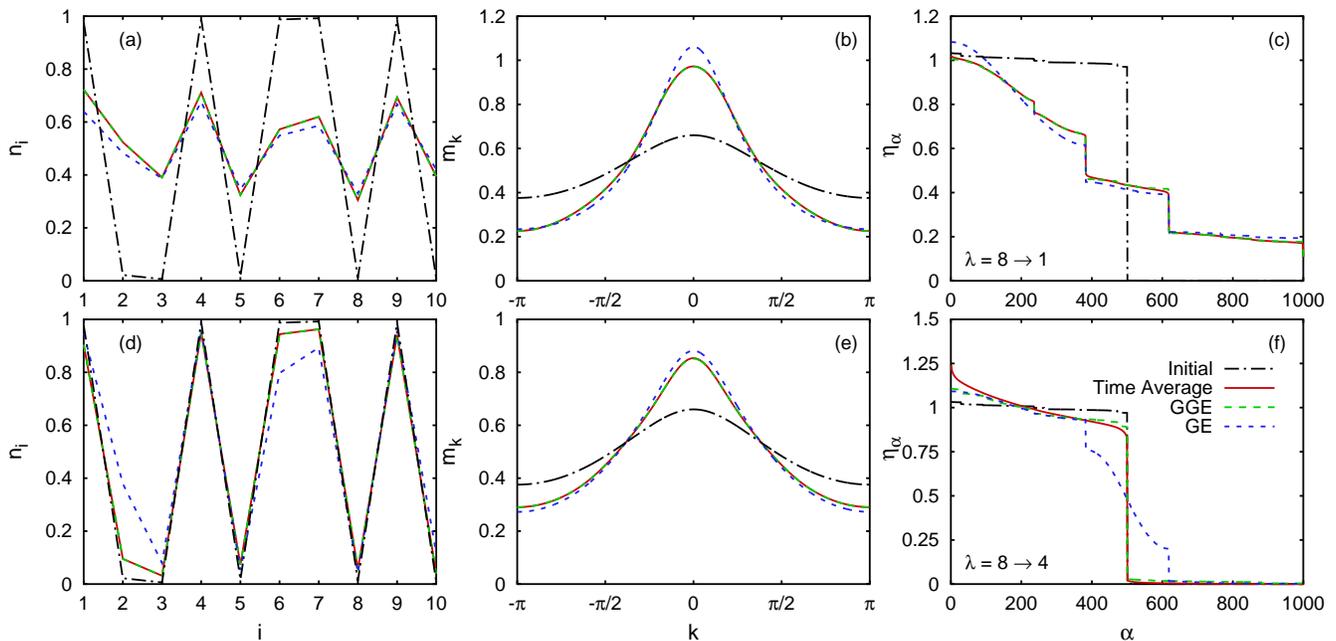}
	\caption{(Color online) As Fig. \ref{fig:lambda=0->1,4} but for quenches from 
$\lambda_I=8$, i.e., from deep inside the Bose-glass phase.}
	\label{fig:lambda=8->1,4}
\end{figure*}

\twocolumngrid

Two other identifying properties of the initial state, which signal the existence of off-diagonal 
quasi-long range correlations, are the presence of a sharp peak in $m_k$ at $k=0$ 
[Figs.~\ref{fig:lambda=0->1,4}(b) and \ref{fig:lambda=0->1,4}(e)] and in $\eta_\alpha$ at $\alpha=0$ 
[Figs.~\ref{fig:lambda=0->1,4}(c) and \ref{fig:lambda=0->1,4}(f)]. The quenches can be seen to lead 
to a dramatic decrease of the height of those peaks after relaxation, which is similar to the effect of 
finite temperature in equilibrium systems \cite{rigol_05_19}. For $m_k$ and $\eta_\alpha$, a stark 
contrast can be observed between the results obtained for the quench $\lambda_I=0\rightarrow\lambda_F=1$ 
and those obtained for the quench $\lambda_I=0\rightarrow\lambda_F=4$. While, in the former, the 
time-averaged results and the GGE predictions are almost indistinguishable from each other, the same 
is not true for the latter. This suggests that the transition to localization plays an important 
role in the description after relaxation. In addition, the thermal values for both observables in the 
GE are clearly different from the results after relaxation.

Qualitatively, we have obtained a very similar picture to the one gained through 
Fig.~\ref{fig:lambda=0->1,4}, for what happens after relaxation in the extended and localized regimes, 
for a wide range of different initial states. Among those, we considered ground and excited states 
of hard-core-boson Hamiltonians in the form of Eq.~\eqref{eq:Hamiltonian} but with different local 
potentials, including period-2 superlattices 
\cite{rigol_muramatsu_06_26,rigol_fitzpatrick_61,he_rigol_12}. In Fig.~\ref{fig:lambda=8->1,4}, we show 
results for the case in which the initial state is the ground state of Eq.~\eqref{eq:Hamiltonian} with 
$\lambda=8$. In contrast to the case with $\lambda_I=0$, for $\lambda_I=8$ the initial state is deep inside
the Bose-glass phase where the density is inhomogeneous [Figs.~\ref{fig:lambda=8->1,4}(a) and 
\ref{fig:lambda=8->1,4}(d)] and the system lacks coherence. The latter is reflected by the almost flat 
initial momentum distribution [Figs.~\ref{fig:lambda=8->1,4}(b) and \ref{fig:lambda=8->1,4}(e)]. 
Localization in this regime is revealed by the natural orbital occupations [Figs.~\ref{fig:lambda=8->1,4}(c) 
and \ref{fig:lambda=8->1,4}(f)], which is nearly 1 for the first 500 orbitals (there are
500 particles in the system), i.e., the bosons in this many-body system can be seen as single 
particles localized within a few sites. This picture is confirmed by the form of the natural orbital wave 
functions (not shown).

After the relaxation dynamics following the quenches $\lambda_I=8\rightarrow\lambda_F=1$ and 
$\lambda_I=8\rightarrow\lambda_F=4$, one can infer from Fig.~\ref{fig:lambda=8->1,4} 
[panels (b), (c), (e), and (f)] that one-particle correlations are enhanced from the ones in the 
initial state. This follows as the height of the zero-momentum occupations increases, the zero-momentum
peaks become narrower, and the occupation of the lowest natural orbitals depart from 1. 
This is very different from what happens in the quenches $\lambda_I=0\rightarrow\lambda_F\neq0$ 
depicted in Fig.~\ref{fig:lambda=0->1,4}, where one-particle correlations are reduced. Despite 
this contrast, we find that the GGE results are almost indistinguishable from the time-averaged ones 
for all observables in quenches $\lambda_I=8\rightarrow\lambda_F=1$, while for quenches 
$\lambda_I=8\rightarrow\lambda_F=4$ only the density and $m_k$ are accurately described by the GGE. 
In the latter quench, the GGE fails to describe the natural orbital occupations, pointing once again 
towards the role of localization.

\subsection*{Scaling with system size}

\begin{figure}[!t]
	\centering
	\includegraphics[width=0.48\textwidth]{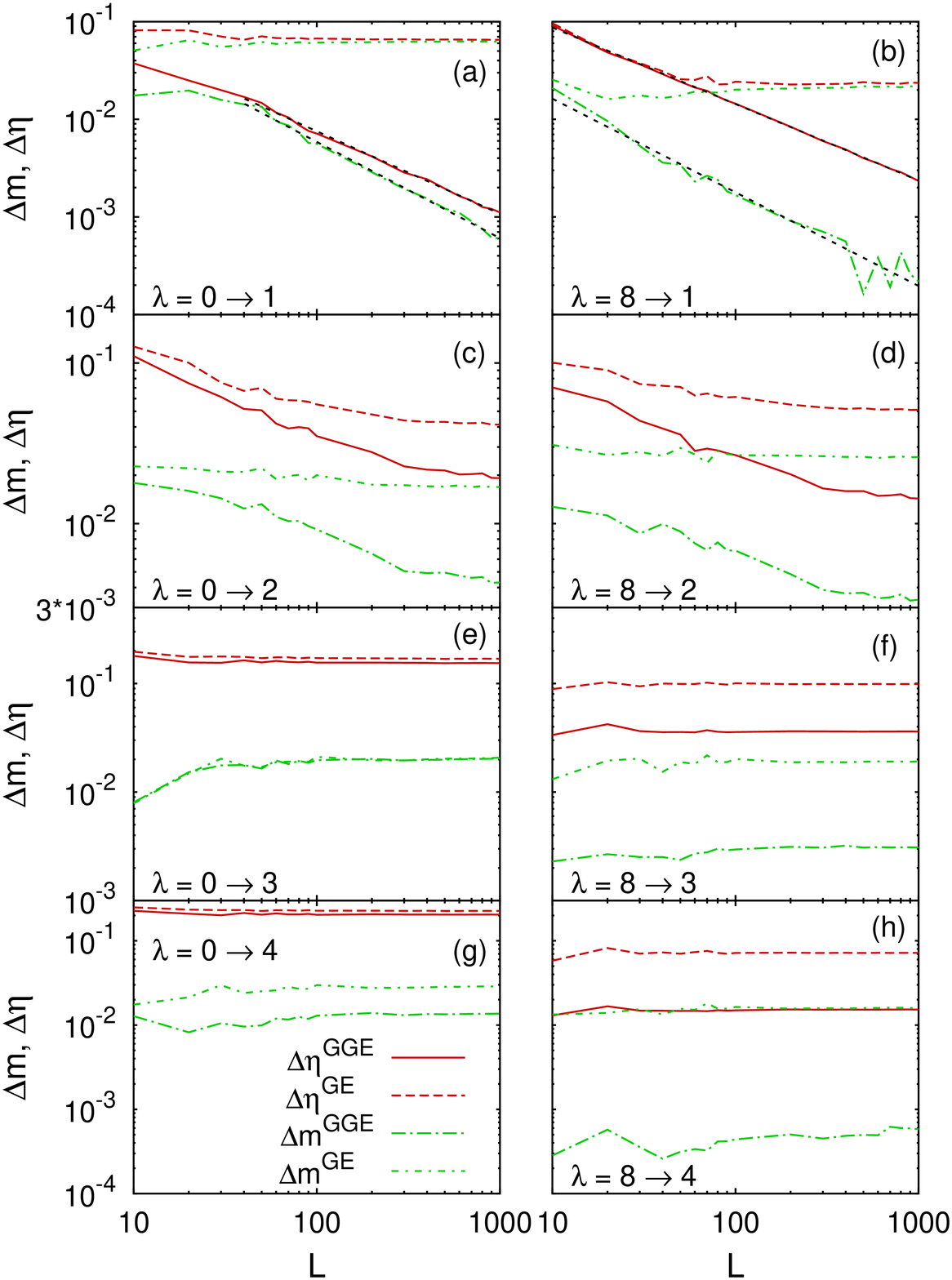}
	\caption{(Color online) Finite-size scaling of $\Delta m^\text{GGE(GE)}$ and 
$\Delta \eta^\text{GGE(GE)}$ for the quenches studied in Figs.~\ref{fig:relaxnk} and \ref{fig:relaxnk2}. 
The dashed lines in (a)--(e) are power-law fits leading to $\Delta m^\text{GGE}\propto L^{-0.99}$ and 
$\Delta \eta^\text{GGE}\propto L^{-0.85}$ in (a), and $\Delta m^\text{GGE}\propto L^{-0.96}$ and 
$\Delta \eta^\text{GGE}\propto L^{-0.78}$ in (b). Up to $100$ sites, the time average was 
taken over $900$ steps with $\tau\in[10^4,10^5]$. For all other system sizes, the time average was taken 
over $437$ steps with $\tau\in[10^4,5.37\times 10^4]$.}
	\label{fig:Scaling}
\end{figure}

Even more important than the actual differences seen in Figs.~\ref{fig:lambda=0->1,4} and 
\ref{fig:lambda=8->1,4} between the long-time averages and the predictions of statistical 
ensembles (GE and GGE) is how those differences scale with increasing system size ($L=1000$ in those 
figures). One could imagine, for example, that while the differences between the time averages and the 
GE are large for finite systems they may disappear in the thermodynamic limit. Another possibility is that 
the differences between the time averages and the GGE are small for the quenches and system sizes 
studied here but they may not vanish in the thermodynamic limit, which would invalidate the GGE 
description for thermodynamic systems. Cases in which integrable systems seemed to behave thermally, 
but failed to exhibit the required scaling with system size, were recently studied in 
Refs.~\cite{rigol_fitzpatrick_61,he_rigol_12}.

In order to study the scaling of the discrepancies between the time averages and the statistical 
predictions, we compute the normalized differences $\Delta O$ between the long-time average 
of the observables $O_i^\text{av}$ and the ensemble predictions $O_i^\text{GGE(GE)}$
\begin{equation}
	\Delta O^\text{GGE(GE)}=\frac{\sum_j \abs{O_j^\text{av}-O_j^\text{GGE(GE)}}}{\sum_j O_j}.
	\label{eq:DeltaGGEnk}
\end{equation}
Note that $O$ stands for $n,\ m$, and $\eta$, and $j$ is a dummy variable that stands for $i$, $k$, 
and $\alpha$, respectively. This quantity is defined in the same spirit as $\delta O$ in 
Eq.~\eqref{eq:deltaO}.

In Fig.~\ref{fig:Scaling}, we show the scaling of $\Delta m^\text{GGE(GE)}$ and $\Delta \eta^\text{GGE(GE)}$ 
for the quenches studied in Figs.~\ref{fig:relaxnk} and \ref{fig:relaxnk2}. Apparent differences can be 
seen between the scalings when $\lambda_F$ lies in the extended, critical, and localized regimes. Different 
initial states, on the other hand, lead to qualitatively similar behavior of $\Delta O^\text{GGE(GE)}$, 
i.e., $\lambda_F$ is the parameter that determines how the outcome of the relaxation dynamics compares 
to the predictions of statistical ensembles.

In quenches terminating in the extended phase [$\lambda_F=1$, Figs.~\ref{fig:Scaling}(a) and 
\ref{fig:Scaling}(b)], one can see that $\Delta m^\text{GGE}$ and $\Delta \eta^\text{GGE}$ exhibit a 
power-law decrease with increasing system size. The small oscillations in $\Delta m^\text{GGE}$, seen 
in Fig.~\ref{fig:Scaling}(b) for the largest system sizes, are due to the small values of this quantity. 
They depend on the exact time intervals and number of time steps used in the time averages. Hence, such 
oscillations are an artifact of our numerical calculations and are not expected to be present if one 
takes the infinite time averages used in previous works \cite{rigol_dunjko_08_34,cassidy_clark_11_56}, 
which are not available here. $\Delta m^\text{GE}$ and $\Delta \eta^\text{GE}$, on the other hand, 
exhibit a clear saturation to finite values with increasing system size. From these scalings, we conclude 
that the GGE correctly describes $m_k$ and $\eta_\alpha$ after relaxation, despite the absence of 
translational invariance and the presence of disorder. On the contrary, the GE fails to describe those 
observables, which makes evident that these systems do not thermalize in the traditional sense.

Quenches terminating at the critical point [$\lambda_F=2$, Figs.~\ref{fig:Scaling}(c) and 
\ref{fig:Scaling}(d)], and except for the largest system sizes, display a behavior that is qualitatively 
similar to the one seen in quenches to the extended regime. Namely, they exhibit a power-law-like decrease 
of $\Delta m^\text{GGE}$ and $\Delta \eta^\text{GGE}$ with increasing system size. However, a tendency 
towards saturation can also be seen in the differences for the largest system sizes. These can be 
attributed to the failure of the observables to relax to stationary values for the times considered 
here (see Figs.~\ref{fig:relaxnk} and \ref{fig:relaxnk2}). Hence, as long as relaxation is achieved, the 
GGE provides a good description of observables also at the critical point. The GE, on the other hand, 
fails to describe $m_k$ and $\eta_\alpha$ (as it does in the extended regime).

The quenches to the localized phase [$\lambda_F=3$, Figs.~\ref{fig:Scaling}(e) and \ref{fig:Scaling}(f), 
and $\lambda_F=4$, Figs.~\ref{fig:Scaling}(g) and \ref{fig:Scaling}(h)] exhibit a very different scaling 
of $\Delta m^\text{GGE}$ and $\Delta \eta^\text{GGE}$ from that observed in those to the extended regime 
and the critical point. One can see in the corresponding panels in Fig.~\ref{fig:Scaling} that, for 
$\lambda_F=3$ and $\lambda_F=4$, $\Delta m^\text{GGE}$ and $\Delta \eta^\text{GGE}$ are almost constant 
with increasing system size, in the same way (up to an offset) that $\Delta m^\text{GE}$ and 
$\Delta \eta^\text{GE}$ are. This makes evident that the GGE description breaks down in the localized 
phase, in a similar way that standard statistical ensembles fail, in general, to describe integrable 
systems after relaxation. We should note, however, that the GGE predictions are closer to the long-time 
averages than the ones provided by the GE, as expected given the larger number of constraints 
imposed in the former ensemble.

\begin{figure}[!t]
  \includegraphics[width=0.48\textwidth]{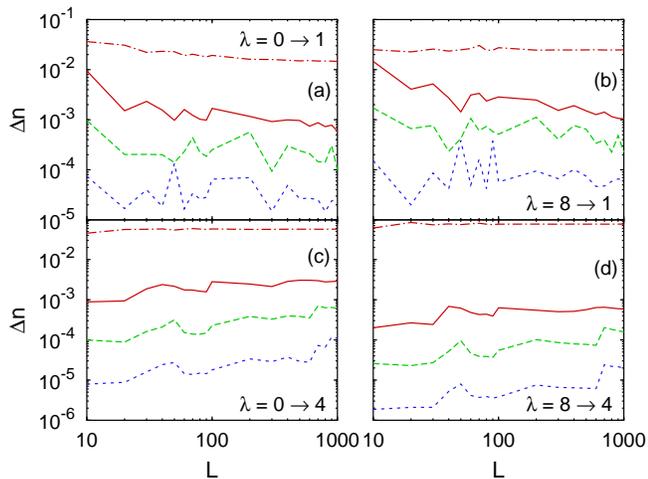}
   \caption{(Color online) Scaling of $\Delta n^\text{GGE}$ and $\Delta n^\text{GE}$ with increasing 
system size. Results for $\Delta n^\text{GGE}$ are reported for time averages calculated using different 
numbers of time steps. Since for the infinite time average $\Delta n^\text{GGE}=0$, the number of time 
steps used in the finite average determines the result. The continuous (red) line shows an average over 99 steps 
with $\tau\in[9.9\times10^5,10^6]$, the dashed (green) line an average over 990 steps with 
$\tau\in[9\times10^5,10^6]$, and the dotted (blue) line an average over 9900 steps with 
$\tau\in[10^4,10^6]$. The dash-dotted (red) line shows $\Delta n^\text{GE}$ for an average over 9900 steps.}
\label{fig:scaling3}
\end{figure}

We have also studied the scaling of the differences $\Delta n^\text{GGE}$ and $\Delta n^\text{GE}$ for 
all parameter regimes depicted in Fig.~\ref{fig:Scaling}. We find that $\Delta n^\text{GE}$ behaves 
similarly to $\Delta m^\text{GE}$ and $\Delta \eta^\text{GE}$, i.e., it saturates to finite values 
with increasing system size. On the contrary, $\Delta n^\text{GGE}$ exhibits a qualitatively 
different behavior from $\Delta m^\text{GGE}$ and $\Delta \eta^\text{GGE}$. 
Independently of $\lambda_F$, we find that $\Delta n^\text{GGE}$ is very small and almost size 
independent. This can be understood because $n_i$ is a property that is shared by hard-
core bosons and noninteracting fermions, and, by construction, the infinite time average of
one-body fermionic observables is given by the GGE. Since for the infinite time average
$\Delta n^\text{GGE}=0$, this quantity is strongly affected by the width of the time interval used to 
calculate the time averages as well as by the number of time steps used. Evidence of this dependence 
is presented in Fig.~\ref{fig:scaling3} for quenches with $\lambda_F=1$ and $\lambda_F=4$ (the results 
for other values of $\lambda_F$ are qualitatively similar).

\subsection*{One-particle correlations}

The three observables we have studied throughout this work provide complementary information about 
one-particle correlations. Two of those observables ($n_i$ and $m_k$) are currently 
accessible in ultracold-gas experiments. In order to conclude our study, and to make contact with the discussion in 
Ref.~\cite{caneva_canovi_11}, we also directly analyze the behavior of one-particle correlations. 
Note that $\rho_{ij}$ is a complex Hermitian matrix, and this is why $n_i$, $m_k$, and $\eta_\alpha$ 
are all real quantities.

\begin{figure}[!t]
  \includegraphics[width=0.48\textwidth]{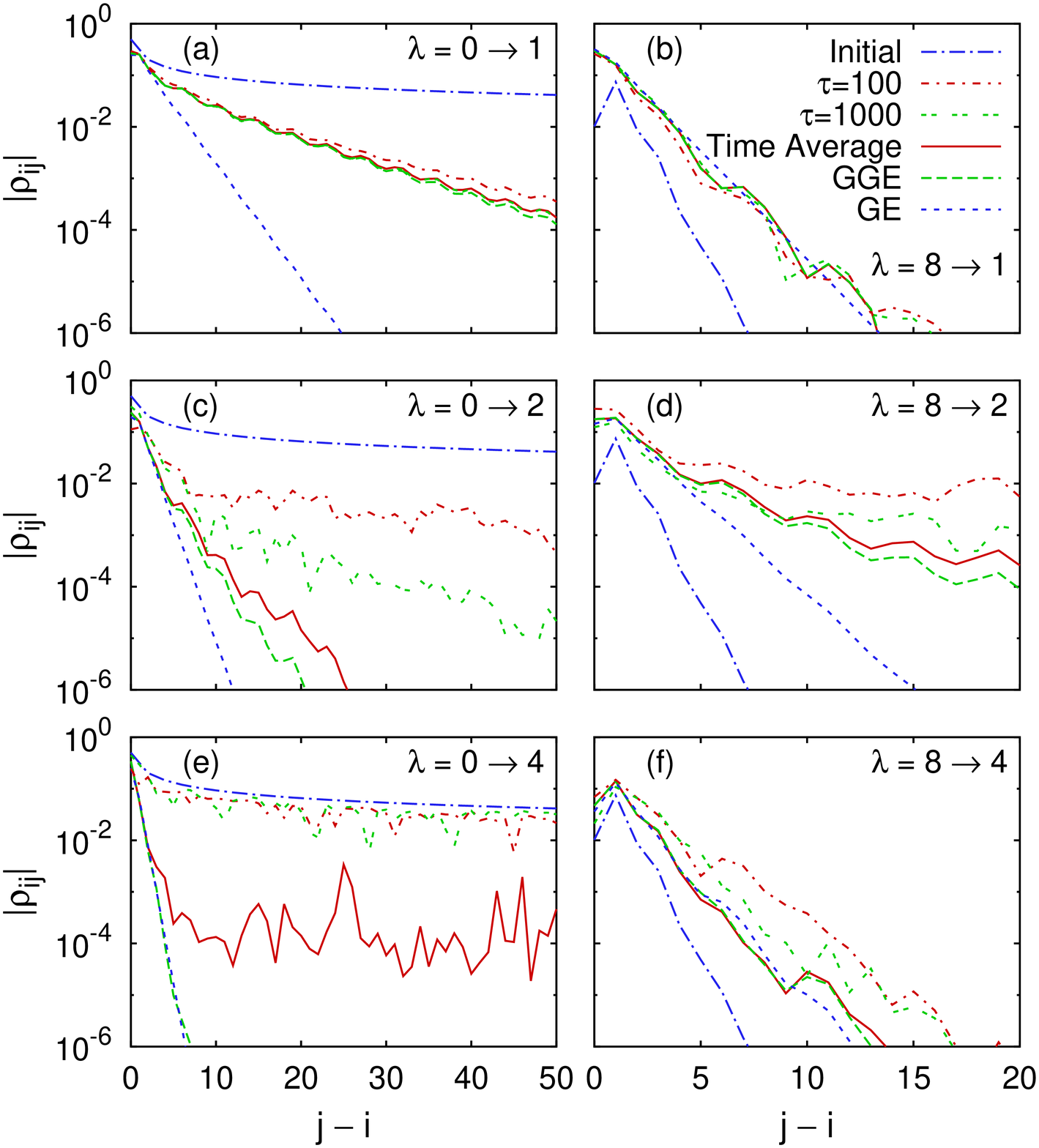}
   \caption{(Color online) Decay of the absolute value of $\rho_{ij}$ for $i=500$ and $j\geq 500$ in a 
system with $L=1000$. The time average was taken over 437 steps with $\tau\in[10^4,5.37\times10^4]$. 
The results depicted are the absolute values after taking those time 
averages [$\rho_{ij}(\tau)$ is complex].}
	 \label{fig:corr1}
\end{figure}

In Fig.~\ref{fig:corr1}, we show how the absolute value of $\rho_{ij}$ decays when $i$ is fixed to be 
the central site in the lattice ($i=L/2$) and $j$ moves towards the boundaries. Results are presented 
for two different initial states for quenches towards the extended, critical, and localized regimes, 
for different times (as well as for the time average), and within the GGE and the GE. The behavior of 
$\rho_{ij}$ in the initial state (in equilibrium $\rho_{ij}$ is real) reflects the nature of the 
ground state in the extended and localized phases. In the former, one-particle correlations exhibit a 
power-law decay ($\rho_{ij}\propto1/\sqrt{|i-j|}$), no matter the value of $\lambda$, while in the 
latter they decay exponentially \cite{he_satija_12_67}.

The quenches towards the extended phase [Figs.~\ref{fig:corr1}(a) and \ref{fig:corr1}(b)] exhibit clear
similarities no matter the value of $\lambda_I$. We find the following: (i) $|\rho_{ij}|$ is very similar, but not 
the same, for $\tau=100$, $\tau=1000$, and the time average. (ii) The time average and the GGE results show 
an excellent agreement with each other. (iii) $\rho_{ij}$ exhibits a faster, and featureless, exponential 
decay in the GE. This is all consistent with our conclusion that the GGE provides an adequate description 
of one-particle observables after relaxation, while the GE fails to do so in this regime. 

Figures \ref{fig:corr1}(c) and \ref{fig:corr1}(d) depict results for quenches to the critical point. In 
this case, due to the slow relaxation dynamics discussed before, the values of $|\rho_{ij}|$ at different 
times differ from each other and from the time average. The time-averaged results can be seen to be 
closest to the GGE prediction and are clearly distinct from those in the GE. Calculating the time 
averages for later times (not depicted) does improve the agreement between those averages and the GGE 
predictions, revealing a picture similar to the one obtained for quenches to the extended phase in 
Figs.~\ref{fig:corr1}(a) and \ref{fig:corr1}(b).

Results for quenches to the localized phase are presented in Figs.~\ref{fig:corr1}(e) and 
\ref{fig:corr1}(f). Once again, $|\rho_{ij}|$ at different times differ from each other and 
from the time average. The latter is also different (although quite close for the quench 
$\lambda_I=8\rightarrow\lambda_F=4$) from the GGE predictions. This is compatible with our previous 
findings that the GGE fails to describe $m_k$ and $\eta_\alpha$ after relaxation in this regime. 
Further understanding of the behavior seen for these quenches can be gained by analyzing the case 
in which $\lambda_F\rightarrow\infty$, so that Hamiltonian \eqref{eq:Hamiltonian} can be written as 
$\Hop=\sum_j \epsilon_j \nbop{j}$, where $\epsilon_j$ is the local chemical potential in each site. 
It then follows that
\begin{equation}
\rho_{ij}(\tau)=\langle \Psi(\tau)\vert b^\dagger_ib^{}_j\vert\Psi(\tau)\rangle\approx
\rho_{ij}(0)e^{i(\epsilon_i-\epsilon_j)\tau},
\end{equation}
which means that if one quenches deep inside the localized phase, $|\rho_{ij}(\tau)|\approx\rho_{ij}(0)$, 
i.e., correlations present in the initial state are preserved, similarly to what we see in 
Fig.~\ref{fig:corr1}(e). 

We note that our results in Figs.~\ref{fig:corr1}(e) and \ref{fig:corr1}(f) are similar to the ones 
reported in Fig.~3 in Ref.~\cite{caneva_canovi_11} for two-point correlations of the order parameter. 
However, the contrasts between Figs.~\ref{fig:corr1}(e) and \ref{fig:corr1}(f) and 
Figs.~\ref{fig:corr1}(a) and \ref{fig:corr1}(b) make evident that the failure of the GGE in disordered 
systems is a consequence of localization and not of the breaking of translational symmetry. Our results 
also make clear the importance of computing time averages, for complex quantities such as $\rho_{ij}$,
before comparing with the predictions of the GGE description.

\section{Summary\label{sec:summary}}

In this work, we studied the dynamics and description after relaxation of hard-core bosons in 
one-dimensional lattices after a sudden change of the strength of an additional quasi-periodic potential. 
This model features two distinct regimes, an extended regime for weak quasi-periodic potentials and a 
localized regime for strong quasi-periodic potentials. Our analysis has shown that the approach of 
observables towards their time-independent values after relaxation comes close to following a power law. For the finite 
system sizes studied, all observables reach their time-independent values within the considered time 
scales. The sole exceptions were the quenches towards the critical point, where the dynamics was found 
to be slower and time-independent values of the observables were not reached for the largest 
lattices. We have argued that, in most of the cases analyzed, the times required for the observables 
to reach their stationary values increase as a power law with the system size.

We further compared the long-time average of observables with statistical descriptions provided 
by the GE and the GGE. The GE failed to describe all observables after relaxation in the quenches considered, 
as expected since these systems are integrable. The GGE, on the other hand, was found to provide a 
good description of observables after relaxation in the extended phase, and at the critical point, whenever 
observables became time independent (up to vanishingly small fluctuations). The scaling behavior in these 
two cases suggests that, in the thermodynamic limit, the GGE results are identical to those after relaxation. 
On the contrary, in the localized regime, we have found that the GGE fails to describe observables that 
depend on nonlocal correlations (such as $m_k$ and $\eta_\alpha$) after relaxation, and that this picture 
does not change with changing system size. The time average of the density, on the other hand, was shown 
to be well described by the GGE in all regimes.

From the outcome of this study, as well as from the results in 
Refs.~\cite{rigol_dunjko_07_27,cassidy_clark_11_56,rigol_muramatsu_06_26,cazalilla_iucci_12}, 
we conclude that localization, and not the breaking of translational symmetry as proposed in 
Ref.~\cite{caneva_canovi_11}, can lead to the breakdown of the GGE description. Our work also poses the 
question of whether modifying the GGE by using a different set of conserved quantities (here we used 
the occupation of the single particle eigenstates of the noninteracting fermionic system to which 
hard-core bosons can be mapped), or adding further conserved quantities, would allow one to describe 
time averages of observables in the localized regime. Recent work on finding optimal 
sets of conserved quantities may shed light on these questions \cite{privatecom}.

\begin{acknowledgments}
This work was supported by NSF under Grant No.~OCI-0904597 and by the Office of Naval Research. 
We thank E. Khatami, K. He and T. Wright for helpful discussions and comments on the manuscript.
\end{acknowledgments}

\end{document}